\documentclass[11pt,twoside]{article}

\usepackage{asp2006}
\usepackage{epsf}
\usepackage{psfig}
\usepackage{lscape}
\usepackage{natbib}
\usepackage{graphicx}

\begin{document}
\title{The dependence of 
some metallicity calibrations of strong-line ratios on nitrogen 
enrichment history}   
\author{Y. C. Liang$^1$,  S. Y. Yin$^{1,2}$}   
\affil{$^1$National Astronomical Observatories, Chinese Academy of Sciences,
  20A Datun Road, Chaoyang District, Beijing 100012, China \\
$^2$Department of Physics, Hebei Normal University,
Shijiazhuang 050016, China; email: ycliang@bao.ac.cn, syyin@bao.ac.cn
 }    

\begin{abstract} 
We show the dependence of the Bayesian and
the N2-derived oxygen abundances on the N-enrichment history of the
star-forming galaxies.  We select 531 metal-poor and $\sim$20,000 metal-rich
galaxies from the SDSS database for this study. Their ``accurate"
O/H abundances are obtained from $T_e$ and $P$-method, respectively.
The discrepancies of the Bayesian and N2-derived abundances 
from these two ``accurate" abundances show obvious correlations
with log(N/O) abundance ratios. 
\end{abstract}

\section{Introduction}   
Metallicity is one of the fundamental parameters to study the property of
galaxy and its evolution history.
The [N~{\sc ii}]/H$\alpha$ strong emission-line ratio in the 
ionized gas in galaxies is a useful tracer to indicate the metallicities
of the galaxies (as N2=log([N~{\sc ii}]/H$\alpha$);
Pettini \& Pagel 2004, PP04; Yin et al. 2006a, Y06a; 
Liang et al. 2006, L06).
 Nitrogen is mainly synthesized in
intermediate- and low-mass stars. 
The N2 index
will therefore become stronger with ongoing star formation
and galaxy evolution until reaching very high metallicities, 
i.e.  12+log(O/H)$\sim$9.0, where [N {\sc ii}] starts to
become weaker due to the very low electron temperature caused from
strong cooling by metal ions. 
The N2 index is not affected very much by dust extinction due to the
close wavelength positions of the two related lines. 
The near infrared spectroscopic instruments
can gather these two lines for the galaxies with
intermediate and high redshifts. 
The N2 index has helped to
provide important information on the metallicities of galaxies at
high-$z$ (Erb et al. 2006). 
However, the resulting log(O/H) abundances from N2 
(and other methods including [N {\sc ii}] line)
may involve the
specific history of N-enrichment in the galaxies.
We check this effect in more details in this paper.

\section{The dependence  on log(N/O) abundance ratios
of the oxygen abundances derived from strong emission-lines} 

The MPA/JHU group used the photoionization models
to simultaneously fit most prominent emission lines 
(including [N {\sc ii}] line), and then
use the Bayesian technique to calculate the likelihood distribution of the
metallicity of each galaxy in the SDSS database
(Tremonti et al. 2004, T04), given as 
(O/H)$_{\rm Bay}$ in this study.
We select 531 (dr4) metal-poor and $\sim$20,000 (dr2) metal-rich
galaxies from the SDSS database for this study. 
We obtain their metallicities of (O/H)$_{T_e}$ and (O/H)$_{\rm N2,P}$
for the metal-poor sample galaxies (Y06a,b), 
and (O/H)$_{\rm N2}$ and (O/H)$_{\rm P05}$ for the metal-rich galaxies
by using the calibrations given by PP04 and Pilyugin \& Thuan (2005, P05),
respectively. 
 The metal-rich sample galaxies have 
 12+log(O/H)$_{\rm Bay}$$>$8.5 and 12+log(O/H)$_{\rm P05}$$>$8.25 (Y06b). 
 Figs.1a-e show clearly the dependence of these abundances
 on log(N/O). 

\begin{figure}
\centering
\psfig{file=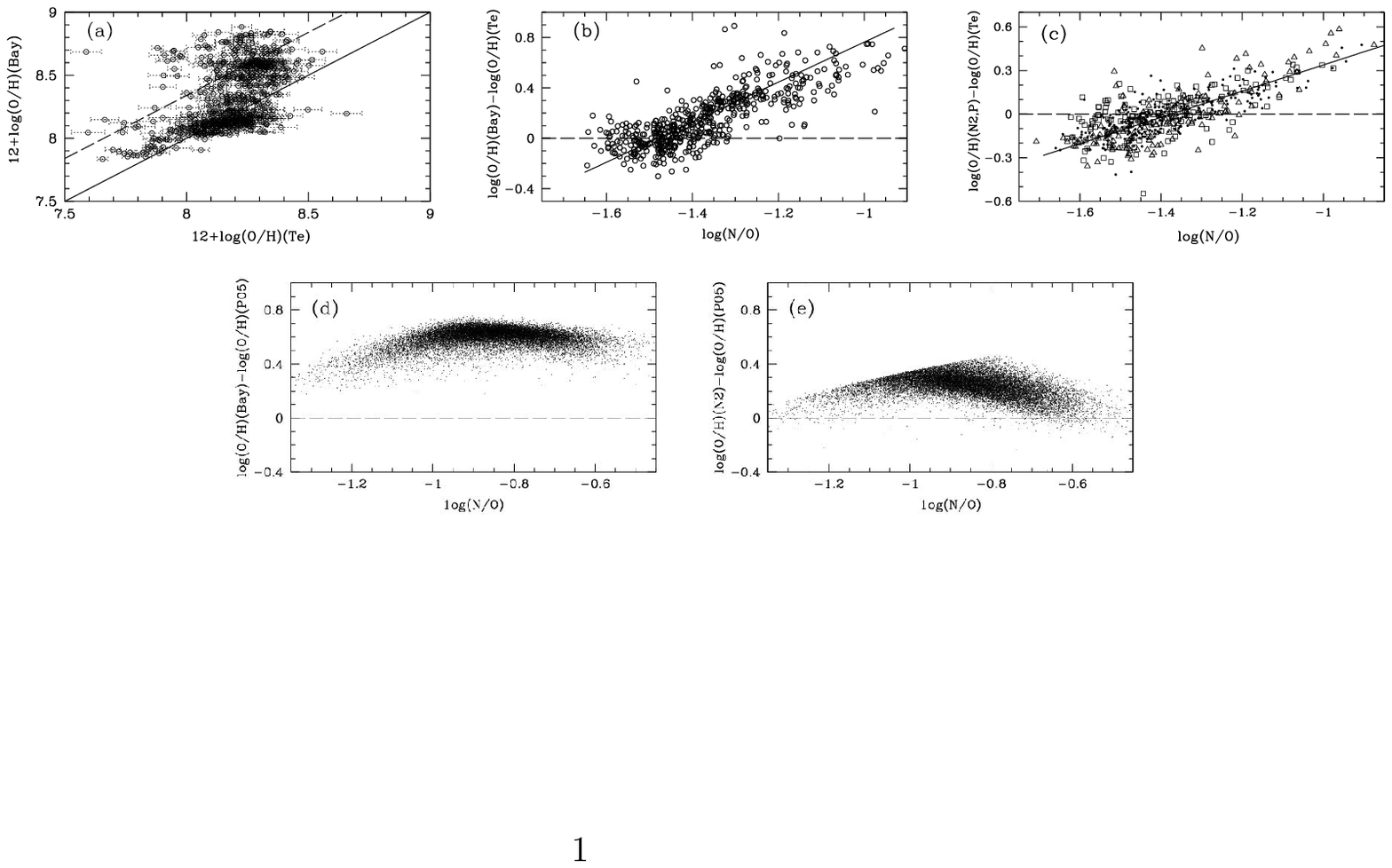,width=11.5cm,angle=0,%
    bbllx=116pt,bblly=208pt,bburx=575pt,bbury=381pt,clip=} 
\caption {{\bf (a)} Comparison between the log(O/H)$_{T_e}$
and log(O/H)$_{\rm Bay}$ for the 531 metal-poor SDSS sample galaxies.
The long-dashed line represents a +0.34\,dex discrepancy from the
equal-value line (the solid line) to show the overestimates of 
the  Bayesian abundances for almost half of the sample galaxies ($\sim$227).
{\bf (b)} The offset between the log(O/H)$_{\rm Bay}$ and 
log(O/H)$_{T_e}$ abundances as a function of the 
log(N/O) abundance ratios for the sample galaxies. 
{\bf (c)} The offset between the log(O/H)$_{\rm N2,P}$ and log(O/H)$_{T_e}$ 
abundances as a function of log(N/O)
for the metal-poor galaxies. See Y06a for more details.
{\bf (d)} The discrepancy between the log(O/H)$_{\rm Bay}$ and 
log(O/H)$_{\rm P05}$ abundances as a function of
log(N/O) for the $\sim$20,000 SDSS metal-rich galaxies,
which shows that, for the galaxies with about log(N/O)$<-0.95$, 
the discrepancy depends on log(N/O) obviously, 
but for those with log(N/O)$\geq -0.95$, the dependence is very weak.
{\bf (e)} The discrepancy between the log(O/H)$_{\rm N2}$ 
(by using PP04's calibration) 
and log(O/H)$_{\rm P05}$ abundances
as a function of log(N/O) for 
the SDSS metal-rich galaxies, which shows two-phase correlations
for the about log(N/O)$<-0.95$ and $\geq -0.95$, respectively.
} 
\end{figure} 

In summary, the strong-line ratio calibrations of metallicities involved
[N~{\sc ii}] emission-line may be related to the N-enrichment history of the
galaxies, which should be kept in mind when compare the resulted abundances 
with others.

\acknowledgements 
This work was supported by the Natural Science Foundation of China
 (NSFC) Foundation under No.10403006,10373005,10433010.

\end{document}